\documentclass[12pt]{article}
\usepackage{a4wide}
\usepackage{epsfig}
\usepackage{amsmath}

\newlength{\absize}
\catcode`@=11
\def\citer{\@ifnextchar [{\@tempswatrue\@citexr}{\@tempswafalse\@citexr[]}}

\def\@citexr[#1]#2{\if@filesw\immediate
  \write\@auxout{\string\citation{#2}}\fi
  \def\@citea{}\@cite{\@for\@citeb:=#2\do
    {\@citea\def\@citea{--\penalty\@m}\@ifundefined
       {b@\@citeb}{{\bf ?}\@warning
       {Citation `\@citeb' on page \thepage \space undefined}}%
\hbox{\csname b@\@citeb\endcsname}}}{#1}}
\catcode`@=12

\begin{document}
  \thispagestyle{empty}
  \pagestyle{empty}
  \renewcommand{\thefootnote}{\fnsymbol{footnote}}
\newpage\normalsize
    \pagestyle{plain}
    \setlength{\baselineskip}{4ex}\par
    \setcounter{footnote}{0}
    \renewcommand{\thefootnote}{\arabic{footnote}}
\newcommand{\preprint}[1]{%
  \begin{flushright}
    \setlength{\baselineskip}{3ex} #1
  \end{flushright}}
\renewcommand{\title}[1]{%
  \begin{center}
    \LARGE #1
  \end{center}\par}
\renewcommand{\author}[1]{%
  \vspace{2ex}
  {\Large
   \begin{center}
     \setlength{\baselineskip}{3ex} #1 \par
   \end{center}}}
\renewcommand{\thanks}[1]{\footnote{#1}}
\begin{flushright}
\end{flushright}
\vskip 0.5cm

\begin{center}
{\large \bf Searching for effects of Spatial Noncommutativity via
a Penning Trap}
\end{center}
\vspace{1cm}
\begin{center}
Jian-Zu Zhang$\;^{\ast}$
\end{center}
\vspace{1cm}
\begin{center}
Institute for Theoretical Physics, East China University of
Science and Technology, Box 316, Shanghai 200237, P. R. China
\end{center}
\vspace{1cm}

\begin{abstract}
The possibility of testing spatial noncommutativity via a Penning
trap is explored. The case of both space-space and
momentum-momentum noncommuting is considered. Spatial
noncommutativity leads to
 the spectrum of the orbital angular
momentum of a Penning trap possessing fractional values, and in
the limits of vanishing kinetic energy and subsequent vanishing
magnetic field, this system has non-trivial dynamics. The dominant
value of the lowest orbital angular momentum is $\hbar/4$, which
is a clear signal of spatial noncommutativity. An experimental
verification of this prediction by a Stern-Gerlach-type experiment
is suggested.
\end{abstract}

\begin{flushleft}
$^{\ast}$ E-mail address: jzzhang@ecust.edu.cn

\end{flushleft}
\clearpage
Studies of low energy effective theory of superstrings show that
space is noncommutative \cite{CDS-SE}. Spatial non-commutativity
is apparent near the Planck scale. Its modifications to ordinary
quantum theory are extremely small. We ask whether one can find
some low energy detectable relics of physics at the Planck scale
by current experiments. Such a possibility is inferred from the
incomplete decoupling between effects at high and low energy
scales. For the purpose of clarifying phenomenological low energy
effects, quantum mechanics in noncommutative space (NCQM) is
available. If NCQM is a realistic physics, all the low energy
quantum phenomena should be reformulated in it. In the literature,
NCQM have been studied in detail \citer{CST,JZZ04c}; many
interesting topics, from the Aharonov-Bohm effect to the quantum
Hall effect have been considered \cite{CDPST00}. Recent
investigation of the non-perturbation aspect of the deformed
Heisenberg-Weyl algebra (the NCQM algebra) in noncommutative space
shows \cite{JZZ04a} that, when the state vector space of identical
bosons is constructed by generalizing one-particle quantum
mechanics, in order to maintain Bose-Einstein statistics at the
non-perturbation level described by deformed annihilation-creation
operators, the consistent ansatz of commutation relations of the
phase space canonical variables should include both space-space
noncommutativity and momentum-momentum noncommutativity. This
explores some new features of effects of spatial noncommutativity
\citer{JZZ04a,JZZ04c}. In \cite{JZZ04b} the possibility of testing
spatial noncommutativity via Rydberg atoms is explored. But the
measurement in experiments of Rydberg atoms depends on a
characteristic frequency which is extremely high and may be
difficult to reach by current experiments. For Penning traps
\citer{Penn,Kost} the situation is different. In this paper we
show a possibility of testing spatial noncommutativity via a
Penning trap. In \cite{JZZ05b} properties of a Penning trap at the
level of NCQM are investigated. Spatial noncommutativity leads to
the spectrum of the orbital angular momentum of a Penning trap
possessing a fractional zero-point angular momentum, fractional
values and fractional intervals. In the limit of vanishing kinetic
energy this system has non-trivial dynamics. We find that in both
limits of vanishing kinetic energy and the subsequent diminishing
magnetic field the dominant value of the lowest orbital angular
momentum in a Penning trap is $\hbar/4$. This result is a clear
signal of spatial noncommutativity, and can be verified by a
Stern-Gerlach-type experiment.

The objects trapped in a Penning trap are charged particles or
ions. The trapping mechanism combines an electrostatic quadrupole
potential $\phi=V_0(-\frac{1}{2}\hat x_i^2+\hat x_3^2)/2d^2$ (i=1,
2) (Henceforth the summation convention is used) and a uniform
magnetic field ${\bf B}$ aligned along the $z$ axis. The
parameters $V_0(>0)$ and $d$ are the characteristic voltage and
length. The particle oscillates harmonically with an axial
frequency $\omega_z=(qV_0/\mu d^2)^{1/2}$ (charge $q>0$) along the
axial direction (the $z$-axis), and in the $x-y$ plane, executes a
superposition of a fast circular cyclotron motion of a cyclotron
frequency $\omega_c=qB/\mu c$ with a small radius and a slow
circular magnetron drift motion of a magnetron frequency
$\omega_m\equiv \omega_z^2/2\omega_c$ in a large orbit. Typically
quadrupole potential $V_0$ superimposed upon the magnetic field
$B$ is a relatively weak addition in the sense that the hierarchy
of frequencies is $\omega_m<<\omega_z<<\omega_c$. The Hamiltonian
$\hat H$ of a Penning trap can be decomposed into a
two-dimensional Hamiltonian $\hat H_2$ and a one-dimensional
harmonic Hamiltonian $\hat H_z$: $\hat H=\hat H_2+\hat H_z$, $\hat
H_z= \frac{1}{2\mu}\hat p_3^2 +\frac{1}{2}\mu\omega_z^2\hat
x_3^2$, and $\hat H_2$ is \citer{Penn,Kost}
\begin{equation}
\label{Eq:H-2} \hat H_2= \frac{1}{2\mu}\hat p_i^2
+\frac{1}{8}\mu\omega_p^2\hat x_i^2
-\frac{1}{2}\omega_c\epsilon_{ij}\hat x_i\hat p_j,
\end{equation}
where $\mu$ is the particle mass, $\omega_p\equiv
\omega_c(1-4\omega_m/\omega_c)$. If NCQM is a realistic physics,
low energy quantum phenomena should be reformulated in this
framework. Thus in the above we formulate the Hamiltonian of the
Penning trap in terms of the deformed canonical variables $(\hat
x_i, \hat p_i)$ of noncommutative space. In order to explore the
new features of such a Penning trap, our attention is focused on
the investigation of $\hat H_2$ and the $z$ component of the
orbital angular momentum $\hat L_z=\epsilon_{ij}\hat x_i\hat p_j$.

In the following we review the background of the NCQM Algebra
\cite{JZZ04a}. In order to develop the NCQM formulation we need to
specify the phase space and the Hilbert space on which operators
act. The Hilbert space can consistently be taken to be exactly the
same as the Hilbert space of the corresponding commutative system
\citer{CST}.

As for the phase space we consider both space-space
noncommutativity (space-time noncommutativity is not considered)
and momentum-momentum noncommutativity. There are different types
of noncommutative theories, for example, see a review paper
\cite{DN}.

In the case of both space-space and momentum-momentum noncommuting
the consistent NCQM algebra \cite{JZZ04a} is:
\begin{equation}
\label{Eq:xp} [\hat x_{I},\hat x_{J}]=i\xi^2\theta_{IJ}, \qquad
[\hat x_{I},\hat p_{J}]=i\hbar\delta_{IJ}, \qquad [\hat p_{I},\hat
p_{J}]=i\xi^2\eta_{IJ},\;(I,J=1,2,3)
\end{equation}
where $\theta_{IJ}$ and $\eta_{IJ}$ are the antisymmetric constant
parameters, independent of the position and momentum. We define
$\theta_{IJ}=\epsilon_{IJK}\theta_K$, where $\epsilon_{IJK}$ is a
three-dimensional antisymmetric unit tensor. We put
$\theta_3=\theta$ and the rest of the $\theta$-components to zero
(which can be done by a rotation of coordinates), then we have
$\theta_{ij}=\epsilon_{ij}\theta$ $(i,j=1,2)$, where
$\epsilon_{ij3}$ is rewritten as a two-dimensional antisymmetric
unit tensor $\epsilon_{ij}$, $\epsilon_{12}=-\epsilon_{21}=1,$
$\epsilon_{11}=\epsilon_{22}=0$. Similarly, we have
$\eta_{ij}=\epsilon_{ij}\eta$. In (\ref{Eq:xp})
$\xi=(1+\theta\eta/4\hbar^2)^{-1/2}$ is the scaling factor which
is necessary for consistent perturbation expansions of $(\hat x_i,
\hat p_j)$ in terms of the undeformed canonical variables $(x_i,
p_j )$ (See Eqs.~(\ref{Eq:hat-x-x})).

We construct the deformed annihilation-creation operators $(\hat
a_i$, $\hat a_i^\dagger)$ which are related to the deformed
canonical variables $(\hat x_i, \hat p_i)$:
$\hat a_i=\sqrt{\frac{\mu\omega_p}{4\hbar}}\left (\hat x_i
+i\frac{2}{\mu\omega_p}\hat p_i\right)$.
When the state vector space of identical bosons is constructed by
generalizing one-particle quantum mechanics, in order to maintain
Bose-Einstein statistics at the deformed level described by $\hat
a_i$ the basic assumption is that operators $\hat a_i$ and $\hat
a_j$ should be commuting. This requirement leads to a consistency
condition of the NCQM algebra \cite{JZZ04a}
\begin{equation}
\label{Eq:cc} \eta=\mu^2\omega_p^2 \theta/4.
\end{equation}
This condition relates the parameters $\eta$ to $\theta$. The
commutation relations of $\hat a_i$ and $\hat a_j^\dagger$ are
$[\hat a_1,\hat a_1^\dagger]=[\hat a_2,\hat a_2^\dagger]=1, [\hat
a_1,\hat a_2]=0;\quad [\hat a_1,\hat a_2^\dagger]
=i\xi^2\mu\omega_p \theta/2\hbar$.
Here, the first three equations are the same boson algebra as the
one in commutative space; The last equation is a new type of boson
commutator, called the correlated boson commutator. It encodes
effects of spatial noncommutativity at the deformed level
described by $(\hat a_i, \hat a_j^\dagger)$, and correlates
different degrees of freedom to each other. It is worth noting
that this new commutator is consistent with {\it all} principles
of quantum mechanics and Bose-Einstein statistics. If
momentum-momentum were commuting, $\eta= 0$, we could not obtain
$[\hat a_i,\hat a_j]=0$. It is clear that in order to maintain
Bose-Einstein statistics for identical bosons at the deformed
level we should consider both space-space noncommutativity and
momentum-momentum noncommutativity. In this paper
momentum-momentum noncommutativity means the {\it intrinsic}
noncommutativity. It differs from the momentum-momentum
noncommutativity in an external magnetic field; In that case the
corresponding noncommutative parameter is determined by the
external magnetic field. Here both parameters $\eta$ and $\theta$
should be extremely small, which is guaranteed by the consistency
condition (\ref{Eq:cc}).

Now we consider perturbation expansions of $(\hat x_i, \hat p_j)$.
The NCQM algebra (\ref{Eq:xp}) has different perturbation
realizations \cite{NP}. We consider the following consistent
ansatz of the perturbation expansions of $\hat x_{i}$ and $\hat
p_{i}$
\begin{equation}
\label{Eq:hat-x-x} \hat
x_{i}=\xi(x_{i}-\theta\epsilon_{ij}p_{j}/2\hbar
 ), \quad \hat
p_{i}=\xi(p_{i}+\eta\epsilon_{ij}x_{j}/2\hbar).
\end{equation}
where $x_{i}$ and $p_{i}$ satisfy the undeformed Heisenberg-Weyl
algebra $[x_{i},x_{j}]=[p_{i},p_{j}]=0,
[x_{i},p_{j}]=i\hbar\delta_{ij}.$

From (\ref{Eq:xp}) and (\ref{Eq:cc}) it follows that $[\hat
L_z,\hat H_2]=0.$ Thus $\hat H_2, \hat L_z$ constitute a complete
set of observables of the two-dimensional sub-system. Using
(\ref{Eq:hat-x-x}) we obtain
\begin{eqnarray}
\label{Eq:CS-H1} \hat H_2=\frac{1}{2M}(
p_i+\frac{1}{2}M\Omega_c\epsilon_{ij} x_j)^2 -\frac{1}{2}K x_i^2
=\frac{1}{2M} p_i^2+\frac{1}{2}\Omega_c\epsilon_{ij} p_i
x_j+\frac{1}{8}M\Omega_p^2 x_i^2,
\end{eqnarray}
where the effective parameters $M, \Omega_c, \Omega_p$ and $K$ are
defined as
$1/M\equiv \xi^2\left(b_1^2/\mu-qV_0
\theta^{\;2}/8d^2\hbar^2\right)$,
$\Omega_c\equiv \xi^2\left(2b_1 b_2/\mu-qV_0
\theta/2d^2\hbar\right)$,
$M\Omega_p^2\equiv \xi^2\left(4b_2^2/\mu-2qV_0/d^2 \right)$,
$K\equiv M(\Omega_c^2-\Omega_p^2)/4$, and
$b_1=1+qB\theta/4c\hbar,\;b_2=qB/2c+\eta/2\hbar.$ The parameter
$K$ consists of the difference of two terms. It is worth noting
that the dominant value of $K$ is $qV_0/2d^2=\mu\omega_z^2/2$,
which is positive.

In the following we are interested in the system  (\ref{Eq:CS-H1})
for the limiting case of vanishing kinetic energy. In this limit
the Hamiltonian (\ref{Eq:CS-H1}) has non-trivial dynamics, and
there are constraints which should be carefully considered
\cite{Baxt,JZZ96,JZZ04b}. For this purpose it is more convenient
to work in the Lagrangian formalism. The limit of vanishing
kinetic energy in the Hamiltonian identifies with the limit of the
mass $M\to 0$  in the Lagrangian \cite{foot1}. The point is that
when the potential is velocity dependent the limit of vanishing
kinetic energy in the Hamiltonian does not corresponds to the
limit of vanishing velocity in the Lagrangian. If the velocity
approached zero in the Lagrangian, there would be no way to define
the corresponding canonical momentum, thus there would be no
dynamics.

The first equation of (\ref{Eq:CS-H1}) shows that in the limit
$M\to 0$ there are constraints
$C_i=p_i+\frac{1}{2}M\Omega_c\epsilon_{ij} x_j=0$,
which should be carefully treated \cite{MZ}. The Poisson brackets
of the constraints 
are $\{C_i,C_j\}_P=M\Omega_c\epsilon_{ij}\ne 0,$ so that the
corresponding Dirac brackets of the canonical variables $x_i, p_j$
can be determined, $\{x_i,p_j\}_D=\delta_{ij}/2,\;
\{x_1,x_2\}_D=-1/M\Omega_c,\;\{p_1,p_2\}_D=-M\Omega_c/4$. The
Dirac brackets of $C_i$  with any variables $x_i$  and $p_j$ are
zero so that the constraints 
are strong conditions, and can be used to eliminate the dependent
variables. If we select $x_1$ and $p_1$ as the independent
variables, from the constraints 
we obtain $x_2=-2p_1/M\Omega_c,\;p_2=M\Omega_cx_1/2.$ We introduce
new canonical variables $q=\sqrt{2}x_1$  and $p=\sqrt{2}p_1$ which
satisfy the Heisenberg quantization condition $[q,p]=i\hbar$, and
define the effective mass $\mu^{\ast}\equiv M^2\Omega_c^2/2K$ and
effective frequency $\omega^{\ast} \equiv K/M\Omega_c$, then the
Hamiltonian $\hat H_2$ reduces to
$H_0=-\left(\frac{1}{2\mu^{\ast}}p^2+
\frac{1}{2}\mu^{\ast}\omega^{\ast2 }q^2\right)$. We define an
annihilation operator
$A= \sqrt{\mu^{\ast}\omega^{\ast}/2\hbar}\;q
+i\sqrt{\hbar/2\mu^{\ast}\omega^{\ast}}\;p,$ and rewrite the
Hamiltonian $H_0$ as
$H_0=-\hbar\omega^{\ast}\left(A^\dagger A+1/2\right)$.
Similarly, we rewrite the angular momentum $\hat L_z$ as
$\hat L_z=\hbar\mathcal{L}^{\ast}\left(A^\dagger A+1/2\right)$,
where
$\mathcal{L}^{\ast}=1-\xi^2\left(M\Omega_c\theta/4\hbar
+\eta/M\Omega_c\hbar\right)$.
The eigenvalues of $H_0$ and $\hat L_z$ are, respectively,
$E_n^{\ast}=-\hbar\omega^{\ast}\left(n+1/2\right)$, 
$\;$
$\mathcal{L}_n^{\ast}=\hbar\mathcal{L}^{\ast}\left(n+1/2\right)$,
$\;$ $(n=0, 1, 2, \cdots)$.
The eigenvalue of $H_0$ is negative, thus unbound. This motion is
unstable. It is worth noting that the dominant value of
$\omega^{\ast}$ is the magnetron frequency $\omega_m$, i.e. in the
limit of vanishing kinetic energy the surviving motion is
magnetron-like, which is more than adequately metastable \cite
{foot2}. The $\theta$ and $\eta$ dependent terms of
$\mathcal{L}^{\ast}$ take fractional values. Thus the angular
momentum possesses fractional eigen values and fractional
intervals. The dominant value, i.e. the $\theta$ and $\eta$
independent term, of the zero-point angular momentum
$\hbar\mathcal{L}^{\ast}/2$ is $\hbar/2$.

For the case of both space-space and momentum-momentum
noncommuting we can consider a further limiting process. After the
sign of $V_0$ is changed, the definition of $\Omega_p$ shows that
the limit of magnetic field $B\to 0$ is meaningful; this implies
that the survived system also has non-trivial dynamics. In this
limit the frequency $\omega_p$ reduces to
$\tilde \omega_p=\sqrt{2}\omega_z$,
the equation (\ref{Eq:cc}) becomes a reduced consistency condition
$\eta=\mu^2\omega_z^2\theta/2$, and
$\xi \to \tilde \xi =(1+\mu^2\omega_z^2\theta^2/8\hbar^2)^{-1/2}$.
The effective parameters $M, \Omega_c, \Omega_p$ and $K$ reduce,
respectively, to the following effective parameters $\tilde M,
\tilde \Omega_c, \tilde \Omega_p$ and $\tilde K$, which are
defined by
$\tilde M \equiv [\tilde \xi^2 \left(1/\mu +
\mu\omega_z^2\theta^2/8\hbar^2\right)]^{-1}= \mu$,
$\tilde \Omega_c\equiv \tilde
\xi^2\left(\eta/\mu\hbar+\mu\omega_z^2\theta/2\hbar\right)=\tilde
\xi^2\mu\omega_z^2\theta/\hbar$,
$\tilde \Omega_p^2\equiv \tilde \xi^2\left(
\eta^2/\mu^2\hbar^2+2\omega_z^2\right)=2\omega_z^2+O(\theta^2)$,
and $\tilde K\equiv \frac{1}{4}\tilde M\left(\tilde
\Omega_c^2-\tilde
\Omega_p^2\right)=-\frac{1}{2}\mu\omega_z^2+O(\theta^2)$.
In this limit $H_0$ and $\hat L_z$ reduce, respectively, to the
following $\tilde H_0$ and $\tilde L_z$:
$\tilde H_0=-\frac{1}{2}\tilde K x_i^2=\hbar\tilde \omega(\tilde
A^\dagger \tilde A+1/2)$,
$\;$ $\tilde L_z=\hbar\mathcal{\tilde L}(\tilde A^\dagger \tilde
A+1/2)$,
$\;$ $\mathcal{\tilde L}=1-\tilde \xi^2(\tilde M\tilde
\Omega_c\theta/4\hbar +\eta/\tilde M\tilde \Omega_c\hbar)$.
Where the annihilation operator
$\tilde A=\sqrt{\tilde \mu\tilde \omega/2\hbar}\;q
+i\sqrt{\hbar/2\tilde \mu\tilde \omega}\;p,$
the effective mass $\tilde \mu\equiv -\tilde M^2\tilde
\Omega_c^2/2\tilde K(>0)$ and the effective frequency $\tilde
\omega^2 \equiv (\tilde K/\tilde M\tilde \Omega_c)^2$.
From the reduced consistency condition
it follows that in $\mathcal{\tilde L}$ the term $\eta/\tilde
M\tilde \Omega_c\hbar=1/2+O(\theta^2)$. The eigenvalues of $\tilde
H_0$ and $\tilde L_z$ are, respectively,
\begin{eqnarray}
\label{Eq:En-Jn}
&&\tilde E_n=\hbar\tilde \omega\left(n+1/2\right), \nonumber\\
&&\mathcal{\tilde L}_n=\hbar\mathcal{\tilde L}\left(n+1/2\right),
\mathcal{\tilde L}=1/2+O(\theta^2), (n=0, 1, 2, \cdots).
\end{eqnarray}
From (\ref{Eq:En-Jn}) we conclude that the dominant value of the
zero-point angular momentum $\hbar\mathcal{\tilde L}/2$ is
$\hbar/4$. This explores the essential new feature of spatial
noncommutativity.

{\bf Testing Spatial Noncommutativity via a Penning Trap} - The
dominant value $\hbar/4$ of the zero-point angular momentum in a
Penning trap can be measured by a Stern-Gerlach-type experiment.
The experiment consists of two parts: the trapping region and the
Stern-Gerlach experimental region. The trapping region serves as a
source of the particles for the Stern-Gerlach experimental region.
After establishing the trap, the experiment includes three steps.
(i) Taking the limit of vanishing kinetic energy. In an
appropriate laser trapping field the speed of the atom can be
slowed to the extent that the kinetic energy term may be removed
\cite {SST}. In the limit of vanishing kinetic energy the harmonic
axial oscillation and the cyclotron motion disappear, only the
magnetron-like motion survives. The magnetron-like motion is
unstable. Fortunately, its damping time is on the order of years
\cite {BG}, so that it is more than adequately metastable. In this
limit, the survived magnetron-like motion slowly drifts in a large
orbit in the $x-y$ plane. At the quantum level, in the limit of
vanishing kinetic energy the mode with the frequency
$\omega^{\ast}$ survives. As we noted before, the dominant value
of $\omega^{\ast}$ is the magnetron frequency $\omega_m$, i.e. the
surviving mode is magnetron-like. (ii) Changing the sign of the
voltage $V_0$ and diminishing the magnetic field $B$ to zero. The
voltage $V_0$ is weak enough so that when the magnetic field $B$
approaches zero the trapped particles can escape along the tangent
direction of the circle from the trapping region and are injected
into the Stern-Gerlach experimental region.
(iii) Measuring the $z$-component of the lowest orbital angular
momentum in the Stern-Gerlach experimental region.
In the case of both space-space
and momentum-momentum noncommuting $L_z$ takes values
$\mathcal{\tilde L}_n$. The dominant value $\hbar/4$ of the
zero-point angular momentum is determined by measuring the
deflection of the beam in the Stern-Gerlach experimental region
\cite{foot3}.

In the case of only space-space noncommuting the dynamics in the
limit of vanishing magnetic field is trivial. The above suggested
experiment can distinguish the case of both space-space and
momentum-momentum noncommuting from the case of only space-space
noncommuting.

For the suggested experiment of Rydberg atoms in \cite{JZZ04b} the
measurement of the lowest angular momentum is frequency dependent,
which in turn  depends on the parameter $\theta$. The magnitude of
$\theta$ is surely extremely small \cite{foot4}; the corresponding
frequency is surely extremely large, which may be difficult to
reach by current experiments. The situation is different in the
Stern-Gerlach-type experiment suggested above, where the
measurement of the lowest angular momentum is frequency
independent.

\vspace{0.4cm}

The author would like to thank Jean-Patrick Connerade for
discussions. This work has been partly supported by the National
Natural Science Foundation of China under the grant number
10074014 and by the Shanghai Education Development Foundation.



\begin{thebibliography}{99}
\bibitem{foot1}
In (\ref{Eq:CS-H1}) in the limit of vanishing kinetic energy,
$\frac{1}{2M}\left( p_i+\frac{1}{2}M\Omega_c\epsilon_{ij}
x_j\right)^2=\frac{1}{2}M \dot{x_i} \dot{x_i}\to 0,$ the
Hamiltonian $\hat H_2$ reduces to $H_0=-\frac{1}{2}K x_i x_i.$ The
Lagrangian corresponding to the Hamiltonian (\ref{Eq:CS-H1}) is
$L=\frac{1}{2}M\dot{x_i}\dot{x_i}
+\frac{1}{2}M\Omega_c\epsilon_{ij}x_i\dot{x_j}+\frac{1}{2}K x_i
x_i$. In the limit of $M\to 0$ this Lagrangian reduces to
$L_0=\frac{1}{2}M\Omega_c\epsilon_{ij}x_i\dot{x_j} +\frac{1}{2}K
x_i x_i.$ From $L_0$ the corresponding canonical momentum is
$p_{0i}=\partial L_0/\partial
\dot{x_i}=\frac{1}{2}M\Omega_c\epsilon_{ji}x_j,$ and the
corresponding Hamiltonian is
$H_0^{\prime}=p_{0i}\dot{x_i}-L_0=-\frac{1}{2}K x_i x_i=H_0.$ Thus
we identify the two limiting processes.
\bibitem{foot2}
The energy in cyclotron motion is almost
exclusively kinetic energy. The energy in harmonic axial
oscillation alternates between kinetic and potential energy.
Reducing the energy in either of these motions reduces their
amplitude. In contrast to these two motions, the energy in the
magnetron-like motion is almost exclusively potential energy. Thus
in the limit of vanishing kinetic energy only the magnetron-like
motion survives. Any process that removes energy from the
magnetron-like motion increases the magnetron radius until the
particle strikes the ring electrode and is lost from the trap.
Fortunately, the damping time $T_m$ for the magnetron-like motion,
for example for the typical electron trap, is on the order of
years \cite {BG}, so that the magnetron-like motion is more than
adequately metastable.
\bibitem{foot3}
A consideration of the spin of the trapped particle changes the
energy spectrum, but does not change the spectrum of the orbital
angular momentum. For the case of spin 1/2, the $z$-component of
the lowest total angular momentum takes dominant values $-\hbar/4$
and $3\hbar/4$. The $z$-component $\hbar/4$ of the lowest orbital
angular momentum is easily red out from the measurement of the
deflections of the beam in the Stern-Gerlach experimental region.
\bibitem{foot4}
There are different bounds on the parameter $\theta$ set by
experiments. The existing experiments on the Lorentz symmetry
violation placed strong bounds on $\theta$ \cite{CHKLO},
$\theta/(\hbar c)^2\le (10 \;TeV)^{-2}$; Measurements of the Lamb
shift \cite{CST} give a weaker bound; Clock-comparison experiments
\cite{MPR} claim a stronger bound. The magnitude of $\theta$ is
surely extremely small.
Measurements of the $\theta$ dependent terms are out of current
experimentat techniques.
\end{thebibliography}
\end{document}